\documentclass[twocolumn]{aastex62}
\usepackage{amsmath}
\graphicspath{{./}{figures/}}

\shorttitle{Searching for Black Holes}
\shortauthors{Siraj \& Loeb}

\begin{document}
\title{Searching for Black Holes in the Outer Solar System with LSST}

\email{amir.siraj@cfa.harvard.edu, aloeb@cfa.harvard.edu}

\author{Amir Siraj}
\affil{Department of Astronomy, Harvard University, 60 Garden Street, Cambridge, MA 02138, USA}

\author{Abraham Loeb}
\affiliation{Department of Astronomy, Harvard University, 60 Garden Street, Cambridge, MA 02138, USA}




\begin{abstract}
Planet Nine has been proposed to potentially be a black hole in the outer solar system. We investigate the accretion flares that would result from impacts of small Oort cloud objects, and find that the upcoming LSST observing program will be able to either rule out or confirm Planet Nine as a black hole within a year. We also find that LSST could rule out or confirm the existence of trapped planet-mass black holes out to the edge of the Oort cloud, indirectly probing the dark matter fraction in subsolar mass black holes and potentially improving upon current limits by orders of magnitude.
\end{abstract}



\section{Introduction}
Observed clustering of extreme trans-Neptunian objects (ETNOs) in the outer solar system suggest the possible existence of a planet with a mass of $\sim 5 - 10 \mathrm{\; M_{\oplus}}$, dubbed Planet Nine, at a distance of $\sim 400 - 800 \mathrm{\; AU}$ from the Sun \citep{2016ApJ...824L..23B, 2019PhR...805....1B}. \cite{2019arXiv190911090S} suggested that Planet Nine could potentially be a black hole (BH) since the likelihood of trapping for a BH may be comparable to that for a free-floating planet. \cite{2020arXiv200400037Z} argued that Planet Nine may not exist, and its observed gravitational effects could potentially be caused by an unobserved ring of small bodies in the outer solar system. There is also the possibility that the clustering is a statistical fluke \citep{2020arXiv200505326C}.

\cite{2017ApJ...834L..20C} proposed the use of interferometry to measure masses of planets from relativistic spacecraft such as those envisioned by Breakthrough Starshot\footnote{https://breakthroughinitiatives.org/initiative/3}. The outer solar system generally \citep{2018AcAau.152..370P} and Planet Nine specifically \citep{Loeb2019} were mentioned as potential targets for Breakthrough Starshot. \cite{2020arXiv200414192W} proposed a search for Planet Nine using sub-relativistic spacecraft, which was further investigated by \cite{2020arXiv200414980L}. \cite{2020arXiv200501120H} showed that the noise due to density and magnetic fluctuations would dominate over Planet Nine's gravitational signal, making such a search infeasible at speeds well above $\sim 10^{-3} \mathrm{\; c} = 300 \mathrm{\; km \; s^{-1}}$. Additionally, such a search would not differentiate between a planet and a BH. Here, we propose a method that does distinguish between a planet and a BH. In particular, we explore the possibility that accretion flares resulting from the tidal disruption of small Oort cloud bodies by a putative Planet Nine BH (PNBH) could power an observable optical signal that could be searched for with the upcoming Vera C. Rubin Observatory Legacy Survey of Space and Time\footnote{https://www.lsst.org/} (LSST). The search method described here is not restricted solely to primordial BHs, but applies generally to subsolar mass BHs including ones produced by other mechanisms \citep{2018PhRvL.120x1102S}.

Our discussion is structured as follows. In Section \ref{sec:td}, we consider the tidal disruption of impactors in the vicinity of a planetary-mass BH, which occurs after impactors are melted due to heating by the Bondi accretion flow of gas in the interstellar medium (ISM). In Section \ref{sec:ir}, we explore the impact rate of small bodies onto a BH in the outer solar system.\footnote{Since there are only $\sim 10^8$ stellar-mass BHs in the Milky Way galaxy \citep{2019arXiv190808775O}, the nearest one should be at a distance $\sim 20 \mathrm{\; pc}$, well beyond the region under consideration here.} In Section \ref{sec:af}, we investigate the accretion flares that would result from such impacts. In Section \ref{sec:ldr}, we compute the rate at which LSST would be expected to detect such accretion flares if a BH in the outer solar system existed. Finally, in Section \ref{sec:d} we summarize key predictions and implications of our model. 


\section{Tidal disruption of impactors}
\label{sec:td}

Since the putative PNBH is located at $400 - 800 \mathrm{\; AU}$ \citep{2019PhR...805....1B}, well beyond the heliopause ($\sim 100 \mathrm{\; AU}$), gas from the ISM will undergo Bondi accretion into the PNBH. For a background ISM density of $\rho_{g, \infty}$ and temperature of $T_{g, \infty}$ (corresponding to a background sound speed of $c_{s, g, \infty} \sim \sqrt{k_B T_{\infty} / m_p}$, where $k_B$ is the Boltzmann constant and $m_p$ is the proton mass) surrounding a PNBH with mass $M_{BH}$, the density and temperature begin to increase significantly beyond their background values interior to a radius of $R_{acc} \sim (2 G M / c_{s, g, \infty}^2)$. In particular, the ISM density \citep{ryden2011dynamics} at a distance from the PNBH, $R \lesssim R_{acc}$, is,

\begin{equation}
    \rho_g \sim \rho_{g, \infty} \left[\frac{G M_{BH} m_p}{2 k_B T_{g, \infty}} \right]^{3/2} R^{-3/2} \; \;.
\end{equation}
Since the ambient gas has pressure while the object is on a ballistic orbit, we assume that the gas encounters the object at a speed comparable to the object's freefall velocity, $v_{ff}$. Given that the freefall velocity for an impactor near the BH is $v_{ff} \sim \sqrt{2 G M_{BH} / R}$, the energy flux from the ISM on the impactor, $\phi_{ff} \sim \rho_g v_{ff}^3$, can be expressed as follows \citep{2012MNRAS.421.1315Z},

\begin{equation}
    \phi_{ff} \sim \rho_{g, \infty} \left[ \frac{(G M_{BH})^2 m_p}{k_B T_{g, \infty}} \right]^{3/2} R^{-3}\; \; .
\end{equation}
Assuming that a large proportion of the energy is re-radiated as blackbody radiation, the surface temperature of the object is $T_{ff} \sim (\phi_{ff} / \sigma_{SB})^{1/4}$, leading to melting within a distance from the PNBH of,

\begin{equation}
    R_{sub} \sim (G M_{BH}) \left(\frac{m_p}{k_B T_{g, \infty} }\right)^{1/2} \left( \frac{\rho_{g, \infty}}{ \sigma_{SB}} \right)^{1/3}  T_{sub}^{-4/3} \; \; ,
\end{equation}
where $\sigma_{SB}$ is the Stefan-Boltzmann constant.
Assuming the impactor is made of water ice, we find its self-consistent sublimation temperature at the ambient ram pressure of $\sim \rho_g v_{ff}^2$ to be $T_{sub} \sim 50 \mathrm{\; K}$ \citep{2006JPCRD..35.1021F}. This yields a sublimation distance,

\begin{equation}
\begin{aligned}
    R_{sub} \sim \; & 6.3 \times 10^6 \mathrm{\; cm} \left( \frac{M_{BH}}{10 \mathrm{\; M_{\oplus}}} \right) \times
    \\ 
    & \left( \frac{\rho_{g, \infty}}{10^{-24} \mathrm{\; g \; cm^{-3}}} \right)^{1/3} \left( \frac{T_{g, \infty}}{10^4 \mathrm{\; K}} \right)^{-1/2} \; \; ,
\end{aligned}
\end{equation}
where the typical background ISM density is $\rho_{g, \infty} \sim 10^{-24} \mathrm{\; g \; cm^{-3}}$ and the typical temperature is $T_{g, \infty} \sim 10^4 \mathrm{\; K}$ \citep{2011piim.book.....D}.

The tidal disruption radius for a self-gravitating body with density $\rho_{obj}$ near a black hole with mass, $M_{BH}$, is,

\begin{equation}
\begin{aligned}
    R_{TD} & \sim \left( \frac{3 M_{BH}}{4 \pi \rho_{obj}} \right)^{1/3}
    \\
    & \sim 10^{9} \mathrm{\; cm} \left( \frac{M_{BH}}{10^{28} \mathrm{\; g}} \right)^{1/3} \left( \frac{\rho_{obj}}{1 \mathrm{\; g \; cm^{-3}}} \right)^{-1/3} \; \; ,
\end{aligned}
\end{equation}
which is well outside of the sublimation radius $R_{sub}$. Since $R_{sub} < R_{TD}$, the sublimation radius, $R_{sub}$, can be considered to be the effective disruption radius for chemically bound icy impactors ($r \lesssim 100 \mathrm{\; m}$, see \citealt{2018ARA&A..56..593W}). The scale of the  Schwarzschild radius does not affect the physics of tidal disruption as long as it is smaller than the tidal disruption radius \citep{2019GReGr..51...30S}.

\section{Impact rate}
\label{sec:ir}

The gravitational focusing factor at a distance $R$ from the BH and a distance $d_{\odot}$ from the Sun is, $\sim (\sqrt{2 G M_{BH} / R} / \sqrt{2 G M_{\odot} / d_{\odot}})^2 = (M_{BH} d_{\odot} / M_{\odot} R)$ (see Appendix A of \citealt{2007MNRAS.375..925J}). For an impactor population described by a power law, $N(>r) \propto r^{1-q}$, with a normalized flux of $F_s$ for objects with radius $r > s$, objects are disrupted by the BH at a rate,

\begin{equation}
    \Gamma \sim \left( \frac{\pi M_{BH} d_{\odot} R F_s}{M_{\odot}}\right) \left( \frac{r}{s} \right)^{-q} \; \; .
\end{equation}
For $R \sim R_{sub}$,

\begin{equation}
\begin{aligned}
    \Gamma \sim & \left( \frac{\pi G M_{BH}^2 d_{\odot} F_s}{M_{\odot}}\right) \left( \frac{\rho_{g, \infty} }{\sigma_{SB} T_m^4}\right)^{1/3} \times 
    \\
    & \left( \frac{m_p }{k_B T_{\infty}} \right)^{1/2} \left( \frac{r}{s} \right)^{-q} \; \; .
\end{aligned}
\end{equation}

For the Kuiper belt, $q \sim 2.6$ for objects sizes in the range of 0.1 - 1 km, and $q \sim 3.7$ for 0.01 - 0.1 km \citep{2012LPICo1667.6348M, 2013AJ....146...36S}. In general, $q \sim 3.5$ for collisionally evolved populations \citep{1969JGR....74.2531D}, which might apply in the limit of small objects.
Since a $\sim 1 \mathrm{\; km}$ object collides with Neptune every $\sim 4 \times 10^3$ years \citep{2003Icar..163..263Z}, $F_{100 \mathrm{m}} \sim 2 \times 10^{-29} \; \mathrm{cm^{-2} \; s^{-2}}$. The space density of Oort cloud objects is nearly uniform in our region of interest \citep{2004come.book..153D, 2019AJ....157..139S}. For sizes smaller than $\sim 100 \mathrm{\; m}$, we consider a single power-law distribution with $q \sim 3.7$, a second option with $q \sim 3.5$, and a third possibility with a transition between the two regimes at $\sim 10 \mathrm{\; m}$. Next, we estimate the parameters of the accretion.

\newpage

\section{Accretion flares}
\label{sec:af}

\begin{figure}[h]
  \centering
  \includegraphics[width=0.9\linewidth]{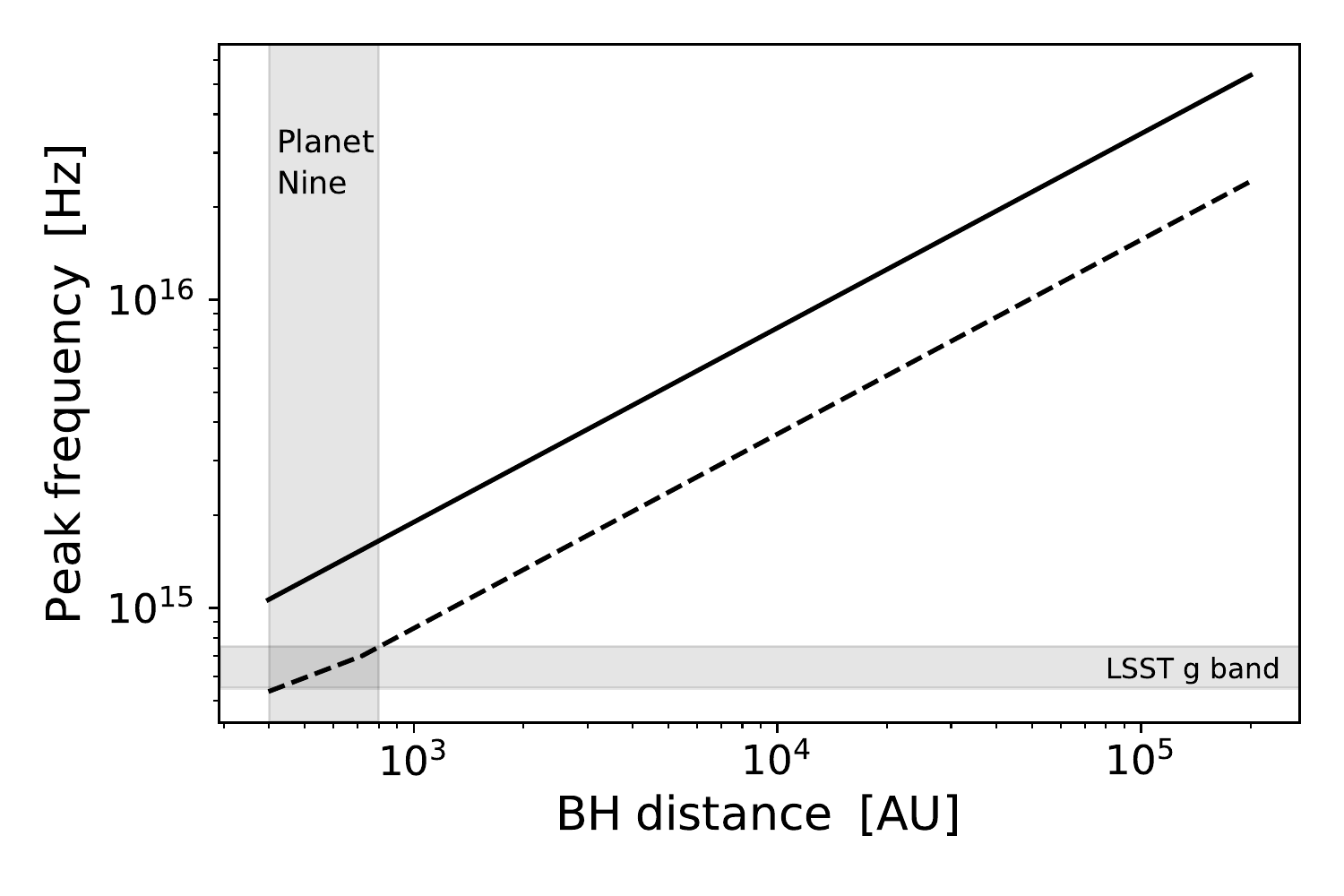}
    \caption{Peak frequency of the flare radiation specrum emitted by an accretion flow as a function of the BH distance from the Sun (in AU), with the range of possible Planet Nine distances and the range of frequencies encompassed by the LSST $g$ band shown for reference. The associated accretion rate $\dot{M}$ corresponds to the LSST flare detection limit, which is discussed in the text. The dotted line corresponds to $M_{BH} \sim 10 M_{\oplus}$ and the solid line to $M_{BH} \sim 5 M_{\oplus}$.
}
    \label{fig:freq}
\end{figure}

The accretion timescale from a radius $R$ is, $\tau_{acc} \sim R^2 / \nu$, where $\nu$ is the kinematic viscosity, $\nu_s \sim \alpha c_s^2 / \Omega_k$, where $\alpha$ is the dimensionless alpha-disk parameter, $c_s$ is the sound speed, and $\Omega_k \sim \sqrt{GM / R^3}$ is the Keplerian angular velocity. In the Advection-Dominated Accretion Flow (ADAF) regime, $c_s \sim 1.2 \times 10^{9} \mathrm{\; cm \; s^{-1}} \left(R/R_S\right)^{-1/2}$, where $R_S$ is the Schwarzschild radius of the BH \citep{2014ARA&A..52..529Y}. As a result, the accretion timescale is,

\begin{equation}
    \tau_{acc} \sim 0.5 \mathrm{\; s} \left( \frac{R}{10^6 \mathrm{\; cm}}\right)^{3/2} \left( \frac{M}{10 \mathrm{\; M_{\oplus}}}\right)^{-1/2} \left( \frac{\alpha}{0.1} \right)^{-1} \; \; .
\end{equation}

Figure 1 of \cite{2014ARA&A..52..529Y} yields a peak frequency of,
\begin{equation}
f_{peak} \sim 10^{15} \mathrm{\; Hz} \left( \frac{M_{BH}}{10 \; M_{\oplus}}\right)^{-0.38} \left( \frac{\dot{M}_{BH} / \dot{M}_{Edd} }{10^{-9}}\right)^{0.36} \; \; ,
\end{equation}
where the Eddington accretion rate $\dot{M}_{Edd} \equiv 10 L_{Edd} / c^2$ is related to the Eddington luminosity $L_{Edd} \equiv 4 \pi G M c /\kappa_{es}$, with $\kappa_{es} \sim 0.4 \mathrm{\; cm^{2} \; g^{-1}}$ being the electron scattering opacity. 

We calculate the accretion rate, $\dot{M}$, during a flare as $\sim M/\tau_{acc}$, where M is the mass of the evaporated impactor interior to $R_{sub}$. The peak frequency as a function of BH distance is shown in Figure \ref{fig:freq}, corresponding to the flare accretion rates at the LSST detection limit, discussed in Section \ref{sec:ldr}. Below and above the peak frequency, the luminosity falls off as a power law with index $\sim 1.3$. For an electron heating parameter of $\delta \sim 0.5$, Figure 2 of \cite{2014ARA&A..52..529Y} yields a radiative efficiency, $\epsilon \equiv L / \dot{M}_{BH} c^2$, of, $\epsilon \sim 6 \times 10^{-5} \; (\dot{M}_{BH} / 10^{-9} \dot{M}_{Edd})^{0.63}$.

\section{LSST detection rate}
\label{sec:ldr}

\begin{figure}
  \centering
  \includegraphics[width=0.9\linewidth]{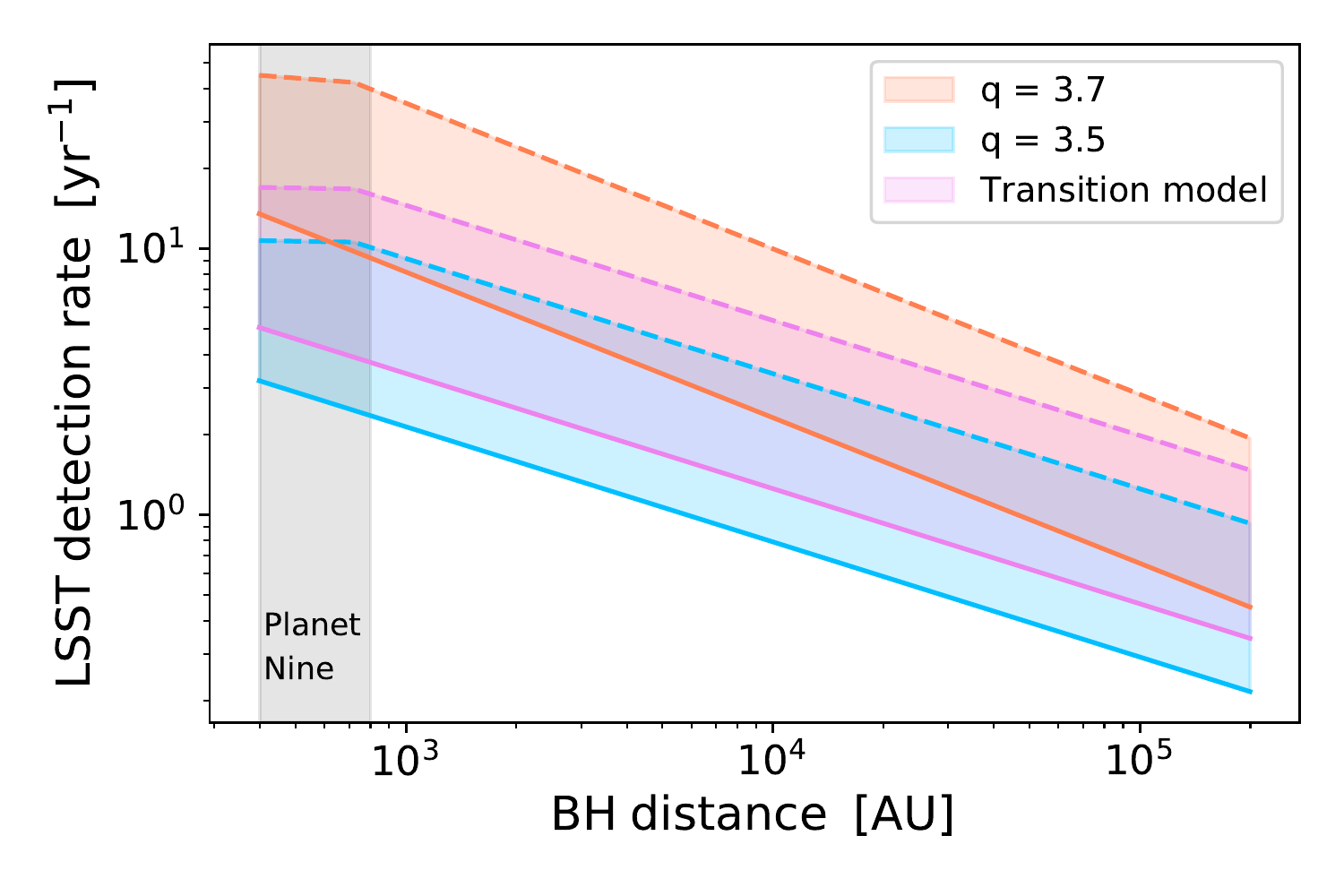}
    \caption{LSST detection rate per year as a function of the BH distance from the Sun (in AU), for $q \sim 3.7$, $q \sim 3.5$, and a broken power-law transition between the two slopes at impactor size $r \sim 10 \mathrm{\; m}$. The range of possible Planet Nine distances are shown for reference. The dotted lines correspond to $M_{BH} \sim 10 M_{\oplus}$ and the solid lines to $M_{BH} \sim 5 M_{\oplus}$.
}
    \label{fig:rate}
\end{figure}

With its field of view of $\sim 9.6 \mathrm{\; deg^2}$ and duty cycle of $\sim 1/3$, LSST will observe $\sim 10^{-4}$ of all flares that originate from point sources and last for a timescale shorter than the exposure time of $\sim 30 \mathrm{\; s}$. LSST's sensitivity in the $g$ band could find a flare with an energy output in the \textit{g} band of $E \sim 10^{19} \mathrm{ \; erg} \; \left( d_{\odot} / 200 \mathrm{\; AU} \right)^2$ over a timescale $\lesssim 30 \mathrm{\; s}$ near the detection limit of $\mathbf{\sim 3 \times 10^{-15} \; \mathrm{erg \; cm^{-2} \; s^{-1}}}$. For peak frequencies above the LSST $g$ band limit, $f_{peak} \gtrsim 7 \times 10^{14} \; \mathrm{Hz}$, we solve the equation, $(4/3) \pi r^3 \rho c^2 \epsilon \nu^{-1.3} = E$, for $r$, and then derive the LSST detection rate, $\sim 10^{-4} \: \Gamma$. The results are shown in Figure \ref{fig:rate}.

\section{Discussion}
\label{sec:d}
We find that if Planet Nine is a BH, its existence can be discovered by LSST due to brief accretion flares powered by small bodies from the Oort cloud, which would be detected at a rate of at least a few per year. Based on the ADAF emission spectrum \citep{2014ARA&A..52..529Y}, such flares would not be expected to have already been discovered by previous optical, X-ray, or radio surveys. If a flare is detected, follow-up integration on the source at a similar flux limit to LSST should yield a flare rate $\sim 10^{4}$ more frequent than observed by LSST, allowing for rapid confirmation of the source as a BH. If multiple bursts are observed over the course of a year, the proper motion of the source can be used to identify the orbital parameters of the BH.

This search method is limited in constraining the subsolar mass BH population since the high speed relative to the Oort cloud would lead to a low disruption rate $\Gamma$, yet it should match the EROS limit of $\sim 10 \%$ of dark matter for PBHs with masses of $\sim 10^{31} \; \mathrm{g}$ \citep{2007A&A...469..387T} by being capable of detecting the nearest one at such a density ($\sim 2 \times 10^5 \mathrm{\; AU}$) at a rate of $\sim 0.3 \mathrm{\; yr^{-1}}$.

However, since the capture rate by the solar system for free-floating planets and BHs with masses of $\sim 5 \mathrm{\; M_{\oplus}}$ may be comparable \citep{2019arXiv190911090S}, and since our method could potentially detect or rule out trapped $\sim 5 \mathrm{\; M_{\oplus}}$ BHs out to a distance of $\sim 10^5 \mathrm{\; AU}$, we could indirectly limit the subsolar mass BH dark matter fraction. Specifically, since the capture rate for a given density scales as the product of cross-section and velocity, and the former scales as $R^2$ while the latter scales as $R^{-1/2}$, in total the capture rate at $\sim 10^5 \mathrm{\; AU}$ would be expected to be a factor of $\sim (10^5 \mathrm{\; AU} / 500 \mathrm{\; AU})^{3/2} \sim 3 \times 10^3$ larger than at the distance of Planet Nine, allowing a non-detection of trapped PBHs over LSST's lifetime to indirectly probe the dark matter fraction of $\sim 5 \mathrm{\; M_{\oplus}}$ BHs to a few times $10^{-5}$, potentially improving on previous limits \citep{2007A&A...469..387T, 2019PhRvD..99h3503N} by orders of magnitude. Additionally, if Planet Nine is a black hole with a magnetic charge, then the synchrotron emission from the accretion flow around it could make its flares much brighter and more easily detectable.

\section*{Acknowledgements}
This work was supported in part by a grant from the Breakthrough Prize Foundation, and Harvard's Black Hole Initiative -- which is funded by grants from JTF and GBMF.





\bibliography{bib}{}

\begin{thebibliography}{}
\expandafter\ifx\csname natexlab\endcsname\relax\def\natexlab#1{#1}\fi
\providecommand{\url}[1]{\href{#1}{#1}}

\bibitem[{{Batygin} {et~al.}(2019){Batygin}, {Adams}, {Brown}, \&
  {Becker}}]{2019PhR...805....1B}
{Batygin}, K., {Adams}, F.~C., {Brown}, M.~E., \& {Becker}, J.~C. 2019,
  \physrep, 805, 1

\bibitem[{{Brown} \& {Batygin}(2016)}]{2016ApJ...824L..23B}
{Brown}, M.~E., \& {Batygin}, K. 2016, \apjl, 824, L23

\bibitem[{{Christian} \& {Loeb}(2017)}]{2017ApJ...834L..20C}
{Christian}, P., \& {Loeb}, A. 2017, \apjl, 834, L20

\bibitem[{{Clement} \& {Kaib}(2020)}]{2020arXiv200505326C}
{Clement}, M.~S., \& {Kaib}, N.~A. 2020, arXiv e-prints, arXiv:2005.05326

\bibitem[{{Dohnanyi}(1969)}]{1969JGR....74.2531D}
{Dohnanyi}, J.~S. 1969, \jgr, 74, 2531

\bibitem[{{Dones} {et~al.}(2004){Dones}, {Weissman}, {Levison}, \&
  {Duncan}}]{2004come.book..153D}
{Dones}, L., {Weissman}, P.~R., {Levison}, H.~F., \& {Duncan}, M.~J. 2004,
  {Oort cloud formation and dynamics}, ed. M.~C. {Festou}, H.~U. {Keller}, \&
  H.~A. {Weaver}, 153

\bibitem[{{Draine}(2011)}]{2011piim.book.....D}
{Draine}, B.~T. 2011, {Physics of the Interstellar and Intergalactic Medium}

\bibitem[{{Feistel} \& {Wagner}(2006)}]{2006JPCRD..35.1021F}
{Feistel}, R., \& {Wagner}, W. 2006, Journal of Physical and Chemical Reference
  Data, 35, 1021

\bibitem[{{Hoang} \& {Loeb}(2020)}]{2020arXiv200501120H}
{Hoang}, T., \& {Loeb}, A. 2020, arXiv e-prints, arXiv:2005.01120

\bibitem[{{Jones} \& {Poole}(2007)}]{2007MNRAS.375..925J}
{Jones}, J., \& {Poole}, L.~M.~G. 2007, \mnras, 375, 925

\bibitem[{{Lawrence} \& {Rogoszinski}(2020)}]{2020arXiv200414980L}
{Lawrence}, S., \& {Rogoszinski}, Z. 2020, arXiv e-prints, arXiv:2004.14980

\bibitem[{{Loeb}(2019)}]{Loeb2019}
{Loeb}, A. 2019, SciTech Europa Quarterly, 31, 1

\bibitem[{{Minton} {et~al.}(2012){Minton}, {Richardson}, {Thomas}, {Kirchoff},
  \& {Schwamb}}]{2012LPICo1667.6348M}
{Minton}, D.~A., {Richardson}, J.~E., {Thomas}, P., {Kirchoff}, M., \&
  {Schwamb}, M.~E. 2012, in Asteroids, Comets, Meteors 2012, Vol. 1667, 6348

\bibitem[{{Niikura} {et~al.}(2019){Niikura}, {Takada}, {Yokoyama}, {Sumi}, \&
  {Masaki}}]{2019PhRvD..99h3503N}
{Niikura}, H., {Takada}, M., {Yokoyama}, S., {Sumi}, T., \& {Masaki}, S. 2019,
  \prd, 99, 083503

\bibitem[{{Olejak} {et~al.}(2019){Olejak}, {Belczynski}, {Bulik}, \&
  {Sobolewska}}]{2019arXiv190808775O}
{Olejak}, A., {Belczynski}, K., {Bulik}, T., \& {Sobolewska}, M. 2019, arXiv
  e-prints, arXiv:1908.08775

\bibitem[{{Parkin}(2018)}]{2018AcAau.152..370P}
{Parkin}, K. L.~G. 2018, Acta Astronautica, 152, 370

\bibitem[{Ryden(2011)}]{ryden2011dynamics}
Ryden, B. 2011, Radiative Gas Dynamics, Department of Astronomy, The Ohio State
  University

\bibitem[{{Schlichting} {et~al.}(2013){Schlichting}, {Fuentes}, \&
  {Trilling}}]{2013AJ....146...36S}
{Schlichting}, H.~E., {Fuentes}, C.~I., \& {Trilling}, D.~E. 2013, \aj, 146, 36

\bibitem[{{Scholtz} \& {Unwin}(2019)}]{2019arXiv190911090S}
{Scholtz}, J., \& {Unwin}, J. 2019, arXiv e-prints, arXiv:1909.11090

\bibitem[{{Shandera} {et~al.}(2018){Shandera}, {Jeong}, \& {Grasshorn
  Gebhardt}}]{2018PhRvL.120x1102S}
{Shandera}, S., {Jeong}, D., \& {Grasshorn Gebhardt}, H.~S. 2018, \prl, 120,
  241102

\bibitem[{{Sheppard} {et~al.}(2019){Sheppard}, {Trujillo}, {Tholen}, \&
  {Kaib}}]{2019AJ....157..139S}
{Sheppard}, S.~S., {Trujillo}, C.~A., {Tholen}, D.~J., \& {Kaib}, N. 2019, \aj,
  157, 139

\bibitem[{{Stone} {et~al.}(2019){Stone}, {Kesden}, {Cheng}, \& {van
  Velzen}}]{2019GReGr..51...30S}
{Stone}, N.~C., {Kesden}, M., {Cheng}, R.~M., \& {van Velzen}, S. 2019, General
  Relativity and Gravitation, 51, 30

\bibitem[{{Tisserand} {et~al.}(2007){Tisserand}, {Le Guillou}, {Afonso},
  {Albert}, {Andersen}, {Ansari}, {Aubourg}, {Bareyre}, {Beaulieu}, {Charlot},
  {Coutures}, {Ferlet}, {Fouqu{\'e}}, {Glicenstein}, {Goldman}, {Gould},
  {Graff}, {Gros}, {Haissinski}, {Hamadache}, {de Kat}, {Lasserre}, {Lesquoy},
  {Loup}, {Magneville}, {Marquette}, {Maurice}, {Maury}, {Milsztajn}, {Moniez},
  {Palanque-Delabrouille}, {Perdereau}, {Rahal}, {Rich}, {Spiro},
  {Vidal-Madjar}, {Vigroux}, {Zylberajch}, \& {EROS-2
  Collaboration}}]{2007A&A...469..387T}
{Tisserand}, P., {Le Guillou}, L., {Afonso}, C., {et~al.} 2007, \aap, 469, 387

\bibitem[{{Walsh}(2018)}]{2018ARA&A..56..593W}
{Walsh}, K.~J. 2018, \araa, 56, 593

\bibitem[{{Witten}(2020)}]{2020arXiv200414192W}
{Witten}, E. 2020, arXiv e-prints, arXiv:2004.14192

\bibitem[{{Yuan} \& {Narayan}(2014)}]{2014ARA&A..52..529Y}
{Yuan}, F., \& {Narayan}, R. 2014, \araa, 52, 529

\bibitem[{{Zahnle} {et~al.}(2003){Zahnle}, {Schenk}, {Levison}, \&
  {Dones}}]{2003Icar..163..263Z}
{Zahnle}, K., {Schenk}, P., {Levison}, H., \& {Dones}, L. 2003, \icarus, 163,
  263

\bibitem[{{Zderic} \& {Madigan}(2020)}]{2020arXiv200400037Z}
{Zderic}, A., \& {Madigan}, A.-M. 2020, arXiv e-prints, arXiv:2004.00037

\bibitem[{{Zubovas} {et~al.}(2012){Zubovas}, {Nayakshin}, \&
  {Markoff}}]{2012MNRAS.421.1315Z}
{Zubovas}, K., {Nayakshin}, S., \& {Markoff}, S. 2012, \mnras, 421, 1315

\end{thebibliography}
\bibliographystyle{aasjournal}



\end{document}